\documentclass{article}
\usepackage[T1]{fontenc} 
\usepackage[utf8]{inputenc} 
\usepackage{ismir,amsmath,cite,url}
\usepackage{graphicx}
\usepackage{color}

\usepackage{soul}
\usepackage{multirow}
\usepackage{booktabs} 
\usepackage{pifont}
\usepackage{enumitem} 
\usepackage{amsfonts} 
\usepackage{enumitem}

\newcommand{\dataset}{Discogs-VI}
\newcommand{\ytdataset}{Discogs-VI-YT}
\newcommand{\model}{Discogs-VINet}
\newcommand{\ncliques}{348,000}
\newcommand{\nversions}{1,900,000}
\newcommand{\ytncliques}{98,000}
\newcommand{\ytnversions}{493,000}
\newcommand{\cmark}{\ding{51}}
\newcommand{\xmark}{\ding{55}}

\setlist{topsep=0pt, leftmargin=*}

\title{Discogs-VI: A Musical Version Identification Dataset Based on Public Editorial Metadata}

\oneauthor
{R. Oguz Araz \hspace{1cm} Xavier Serra \hspace{1cm} Dmitry Bogdanov} 
{Music Technology Group, Universitat Pompeu Fabra, Barcelona\\
{\tt\small \{recepoguz.araz, xavier.serra, dmitry.bogdanov\}@upf.edu}
}

\def\authorname{R. O. Araz, X. Serra, and D. Bogdanov}

\usepackage[bookmarks=false,pdfauthor={\authorname},pdfsubject={\papersubject},hidelinks]{hyperref}

\begin{document}

\maketitle

\begin{abstract} 
Current version identification (VI) datasets often lack sufficient size and musical diversity to train robust neural networks (NNs). Additionally, their non-representative clique size distributions prevent realistic system evaluations. To address these challenges, we explore the untapped potential of the rich editorial metadata in the Discogs music database and create a large dataset of musical versions containing about \nversions{} versions across \ncliques{} cliques. Utilizing a high-precision search algorithm, we map this dataset to official music uploads on YouTube, resulting in a dataset of approximately \ytnversions{} versions across \ytncliques{} cliques. This dataset offers over nine times the number of cliques and over four times the number of versions than existing datasets. We demonstrate the utility of our dataset by training a baseline NN without extensive model complexities or data augmentations, which achieves competitive results on the SHS100K and Da-TACOS datasets. Our dataset, along with the tools used for its creation, the extracted audio features, and a trained model, are all publicly available online.
\end{abstract}

\section{Introduction}\label{sec:introduction}
Artists continue to cover, remix, and reinterpret musical works, creating a rich tapestry of musical versions that celebrate the originals. This proliferation presents a complex challenge: how to accurately identify different versions of a musical work within vast digital catalogs. Version identification (VI) addresses this problem using audio processing methods to find versions of query tracks in music catalogs~\cite{serra_unsupervised_2009, serra_audio_2010, yesiler_less_2020}. VI has thus emerged as a crucial solution with significant implications across multiple applications including music discovery, musicological research, and copyright enforcement. From both the artists' and copyright holders' perspectives, VI has substantial importance as it offers a tool for financial compensation to many music industry stakeholders.\par

Recently, multiple datasets, all derived from scraping the SecondHandSongs\footnote{\url{https://secondhandsongs.com/}} website, were proposed for developing VI systems \cite{xu_key-invariant_2018, yesiler_da-tacos_2019, doras_cover_2019, doras_prototypical_2020}. These datasets have facilitated the development of various systems based on convolutional neural networks (CNNs) \cite{yu_temporal_2019, doras_cover_2019, yesiler_da-tacos_2019, yesiler_accurate_2020, yu_learning_2020, doras_prototypical_2020, du_bytecover_2021, du_bytecover2_2022}. However, their limited sizes have restricted the feasibility of employing larger architectures, such as transformers, which are increasingly utilized in other music information retrieval (MIR) tasks \cite{zeng_training_2023, alonso-jimenez_efficient_2023}. Additionally, existing datasets such as Da-TACOS~\cite{yesiler_da-tacos_2019} and SHS100K~\cite{xu_key-invariant_2018} lack comprehensive metadata, such as genre, style, and release year, which can be useful for detailed performance evaluation and sophisticated training approaches. Furthermore, they fall short in presenting sufficient challenges regarding the distribution of clique sizes, genres, styles, and track durations.\par

This study introduces a significantly larger and more challenging VI dataset. Rather than relying on SecondHandSongs, we use public editorial metadata from the Discogs\footnote{\url{https://www.discogs.com/}} database, which has not been explored in the field previously. Discogs is collaboratively maintained by music enthusiasts and professionals who submit detailed metadata about music releases, including artist details, release information, and extensive credit descriptions. These descriptions not only list track artists and writers but also provide aliases, name variations, and artist relationships, offering a rich framework for identifying versions.\par

Using this metadata, we propose a methodology for identifying a large dataset of versions and mapping this dataset to various music audio collections. The resulting dataset is the largest open-source VI dataset to date. Our contributions can be summarized as follows:\par

\setlist{nolistsep}
\begin{enumerate}
    \item A metadata-only dataset, \dataset{}, containing over \nversions{} versions of around \ncliques{} works.
    \item A subset of this dataset, \ytdataset{}, containing about \ytnversions{} versions of around \ytncliques{} works matched to YouTube URLs of official music uploads. It contains over nine times as many works and over four times as many versions as other datasets.
    \item A larger and more challenging test set that contains other publicly available test sets.
    \item A pre-trained baseline model, \model{}. 
\end{enumerate}

The dataset\footnote{\url{https://mtg.github.io/discogs-vi-dataset/}}, together with the tools for its creation, the extracted audio features, and the model trained on this data\footnote{\url{https://github.com/raraz15/Discogs-VINet}}, are publicly available online.\par

\begin{table*}[!ht]
\setlength{\tabcolsep}{0.115cm}
    \centering
    \small
    \begin{tabular}{@{}llrrrrrcccl@{}}
        \toprule
            Dataset & Source & Cliques & Versions & MCS & ACS & mCS & A-URL & m-URL & OV & Content \\
        \midrule
            covers80 \cite{ellis_covers80_2007} & private & 80 & 160 & 2 & 2 & 2 & - & - & - & Full audio, title, album, artist \\
            YouTubeCovers \cite{silva_music_2015} & YouTube & 50 & 350 & 7 & 7 & 7 & - & - & \xmark & Features (full track) \\
            Da-TACOS \cite{yesiler_da-tacos_2019} & SHS & 1,000 & 13,000 & 13 & 13 & 13 & 1.0 & 1.0 & \xmark & Features (full track), metadata \\
            CoversDataset \cite{doras_cover_2019} & SHS & 26,905 & 110,794 & 24 & 4 & 3 & 1.0 & 1.0 & \xmark & Features (first 3 min) \\
            SHS-100K \cite{xu_key-invariant_2018} & SHS & 9,999 & 116,353 & 387 & 12 & 8 & 1.0 & 1.0 & \xmark & Title, artist \\
        \midrule
            \ytdataset{} & Discogs & 98,785 & 493,049 & 658 & 5 & 2 & 1.5 & 1.0 & \cmark & Rich metadata, features (full track) \\
            \dataset{} & Discogs & 348,796 & 1,911,611 & 1,837 & 6 & 2 & - & - & - & Rich metadata \\
        \bottomrule
    \end{tabular}
    \caption{Overview of publicly-available VI datasets. Da-TACOS refers to the benchmark subset, for which the 2,000 noise works are not reported as they do not form cliques. SHS refers to the SecondHandSongs website. MCS: maximum clique size; ACS: average clique size; mCS: median clique size; A-URL: average YouTube URLs per version; m-URL: median YouTube URLs per version; OV: use of official YouTube videos only. ``-'' denotes that the property is not applicable.}
    \label{tab:datasets}
\end{table*}

\section{Identifying versions on Discogs}\label{sec:discogs}
Discogs database metadata has been previously used in other MIR tasks~\cite{bogdanov_acousticbrainz_2019, bogdanov_quantifying_2017, alonso-jimenez_music_2022}. In this section, we describe the proposed methodology to identify versions and cliques using its metadata. The complete Discogs data is shared as monthly data dumps under a Public Domain license, making it easy to access. In our study, we used the July 2024 data dump.\par

Numerous metadata fields are provided for releases, tracks, and artists, some of which are relevant for VI. We use the track title, track artists, featuring track artists, release artists, track writer artists, and release writer artists metadata. The artist metadata contains unique artist IDs and provides information regarding group memberships, artist aliases, and artist name variations, which we use extensively. In addition, we include genre, style, record label, release format, release date, master release, and release country metadata that can be potentially useful.\par

\subsection{Version finding from metadata}\label{sec:discogs:versions}
We use two critical pieces of information to establish the version relationship between two tracks: the track title and the track writer artists, indicated by the ``Written-By'' metadata field. Specifically, we consider two tracks with the same title and a shared writer artist as versions. This is a sufficient but not necessary condition since two tracks with different names can also be versions. Nonetheless, this condition facilitates finding a significant amount of cliques and versions from the database with high precision.\par

The search for cliques operates on a set of tracks from the database whose track titles are normalized by applying string processing. This includes transliterating Latin characters by removing diacritics, removing leading articles, replacing ``\&'' with ``and'', eliminating any text within parentheses, and removing punctuation marks. These steps aim to mitigate potential differences in metadata between different releases and eliminate mix or edit indicators enclosed in parentheses, e.g., ``(Radio Edit)'', thus facilitating the process of identifying cliques. Later, such differences are considered for differentiating between versions.\par

Using the normalized track titles, we partition the set of tracks into disjoint subsets using exact string matching. Then, we further partition these subsets by the common track writer relation to distinguish different cliques with the same title. To do so, we compile a set of writer artist IDs for every track. Given that an artist on Discogs may represent a group with several members, we extend our collection to contain all associated members and incorporate each artist's known aliases and name variations. As a result of the two-step partitioning, tracks that have the same normalized title and share a track writer are joined in the same cliques. We opted for the shared writer approach because not all writers are consistently included in credits on some releases.\par

Once the cliques are formed, we identify different versions by the track or release artists. In cases where track artists metadata is available, it is used; otherwise, the release artists metadata is used. If there are featuring track artists, they are also included. Therefore, a set of tracks belonging to the same clique and performed by the same set of artists is defined as a version. After identifying the versions, we discard the cliques with only one version.\par

In previous VI datasets, versions are not treated as sets of tracks as in our dataset. This difference arises because Discogs often lists multiple releases for essentially the same version of a track, which may vary only by the year or country of the release. Without direct access to these releases, it is impossible to confirm their differences in advance. Therefore, we treat such tracks as identical versions. Remarkably, our dataset comprehensively includes a variety of version types as systematized in~\cite{serra_identification_2011}, including live versions, remixes, and radio edits, which add valuable diversity and potential utility.\par

The resulting dataset, \dataset{}, contains numerous cliques and versions. Statistics about the dataset in comparison to other datasets are provided in Table~\ref{tab:datasets}.\par

\subsection{Limitations}\label{sec:discogs:limitations}
Due to the complex processes of composing, performing, and releasing music, along with issues related to incomplete or inaccurate metadata, there are potential issues related to our approach.\par

\textbf{Title variability:} Versions can have different names, e.g. ``Moon Over Naples'' is the original version of both ``Spanish Eyes'' and ``Blue Spanish Eyes''. Due to having different names, our algorithm falsely places these versions into different cliques. To address this issue, complementary data from SecondHandSongs or a large language model with music history knowledge can be used.\par

\textbf{Rule-based text matching:} Even for a single language, capturing all syntactic variations with simple rules is difficult. Yet, the database contains many languages with different syntaxes. A music named-entity recognition model may help to resolve this issue.\par

\textbf{Metadata ambiguity:} ``You're My Everything'' is credited to ``Miles Davis'' in some releases while to ``Miles Davis and John Coltrane'', and to ``The Miles Davis Quintet'' in others. These credential differences often arise from practical or legal reasons associated with publishing music. However, we can not know beforehand if they are different versions using only metadata. To reduce duplicate versions, we treat them as the same version.\par

\section{Version Search in YouTube}\label{sec:youtube}
Owing to its detailed metadata, \dataset{} can be mapped to music audio catalogs or other metadata sources. For our research purposes, we use YouTube. To match the Discogs metadata of a version to the YouTube metadata of a video, we design a rule-based algorithm.\par

In the matching process, we only accept videos provided by an official distributor, which can be the artists themselves or third parties such as record labels. This approach is adopted because we expect the official uploads to have more accurate metadata and be more persistent on the platform over time. Consequently, our dataset is the only VI dataset containing official uploads exclusively. In addition, due to this selectivity, our algorithm demonstrates high retrieval accuracy.\par

Discogs provides YouTube URL annotations for some of the releases associated with versions. However, these annotations are not on the track level and they are rarely provided. For a unified approach, we instead query YouTube for all versions. The queries are created using the Discogs version metadata in the format ``artist1, artist2 - track title'', and if featuring artist information is available, we concatenate ``(featuring artist3)''. We then store the top five results for each query and apply our metadata-matching algorithm to all stored results, which allows alternative URLs for certain versions.\par

As a result, we successfully matched 34\% of the versions of \dataset{} to a YouTube URL. Between these matched versions, we were able to download 98\% successfully, corresponding to 33\% of the total versions. We then discarded the versions that were not downloaded and the cliques without at least two downloaded versions to create the \ytdataset{} dataset. It contains 26\% of the versions and 28\% of the cliques of \dataset{}.\par

\subsection{Metadata matching algorithm}\label{sec:youtube:algorithm}
From Discogs metadata, our algorithm utilizes the track title, track artists, or, if unavailable, the release artists, along with any featuring artists. From YouTube, it uses the video's category, uploader, artist, description, duration, and title. We process the strings similarly to the method described in Section \ref{sec:discogs:versions}, except that the punctuation marks and possible texts within parentheses are not deleted to identify different versions.\par

The algorithm initially checks if the video metadata contains the ``Music'' category and if the video is an official YouTube upload. We consider a video official under the following conditions: an artist or a label provided the video, which is indicated in the video description; the video uploader is an artist topic channel auto-generated by YouTube; or the Discogs artist name is the same as the video uploader's. Videos with a duration longer than 20 minutes are discarded to deal with the potential but unlikely issue of tracks sharing their titles with their albums or EPs, which could lead to full-release audio downloads.\par

If a video metadata passes these controls, we use the title and artist information to decide a match. If two titles are equal, we use the artist information. If the titles do not match exactly, we apply some heuristics to strip the video title from any additional information related to remastering, HD, lyrics, etc., and re-attempt the match. We then compare all possible permutations to deal with video titles in the ``artist1, artist2 - track title (featuring artist3)'' format, using exact string matching. This approach makes the dataset less noisy at the cost of losing potential matches.\par

\subsection{Limitations}\label{sec:youtube:limitations}
Since search results and the availability of YouTube videos can be affected by geolocation, re-creating the dataset may yield differences.\footnote{We conducted YouTube queries from Barcelona, Spain.} Moreover, some URLs may become unavailable in the future.\footnote{The URLs were accessed between March 2023 and July 2024.} To mitigate this issue, we provide multiple YouTube URLs per version when possible. Therefore, even if the main URL becomes inactive, numerous versions can still be recovered from alternative URLs. Furthermore, since we only include official uploads, the probability of a video disappearing should be lower than in other datasets. These features have not been considered in previous datasets that share YouTube URLs\cite{xu_key-invariant_2018, lanzendorfer_disco-10m_2024}.\par

Another limitation of our methodology is that less than 8\% of versions are matched to the same YouTube URLs. Analysis showed that almost all of these versions are members of the same cliques. For the cliques that exhibit this issue, we manually kept one of the duplicate versions.\par

\section{Dataset Analysis}\label{sec:dataset}
Following the methodologies described in Section~\ref{sec:discogs} and Section~\ref{sec:youtube}, we created the \dataset{} and \ytdataset{} datasets, respectively. Table~\ref{tab:datasets} reports their sizes. The large amount of detailed metadata in \dataset{} shows great potential: combined with an industrial-scale music audio catalog, it can create new possibilities for VI system development. Moreover, \ytdataset{} contains more clique and version audio than all the others combined, promising to boost model performance and generalization capability.\par

The range of clique sizes in our dataset is unparalleled by others in the field. The presence of cliques with many versions is beneficial for metric learning, as it provides numerous examples within each clique \cite{schroff_facenet_2015}. The average, median, and maximum clique sizes in the dataset indicate that the distribution has a long tail, with the weight concentrated on small clique sizes. Unlike other datasets, this distribution is highly representative of real use cases.\par

\begin{figure}[ht]
    \centering
    \includegraphics[width=\columnwidth]{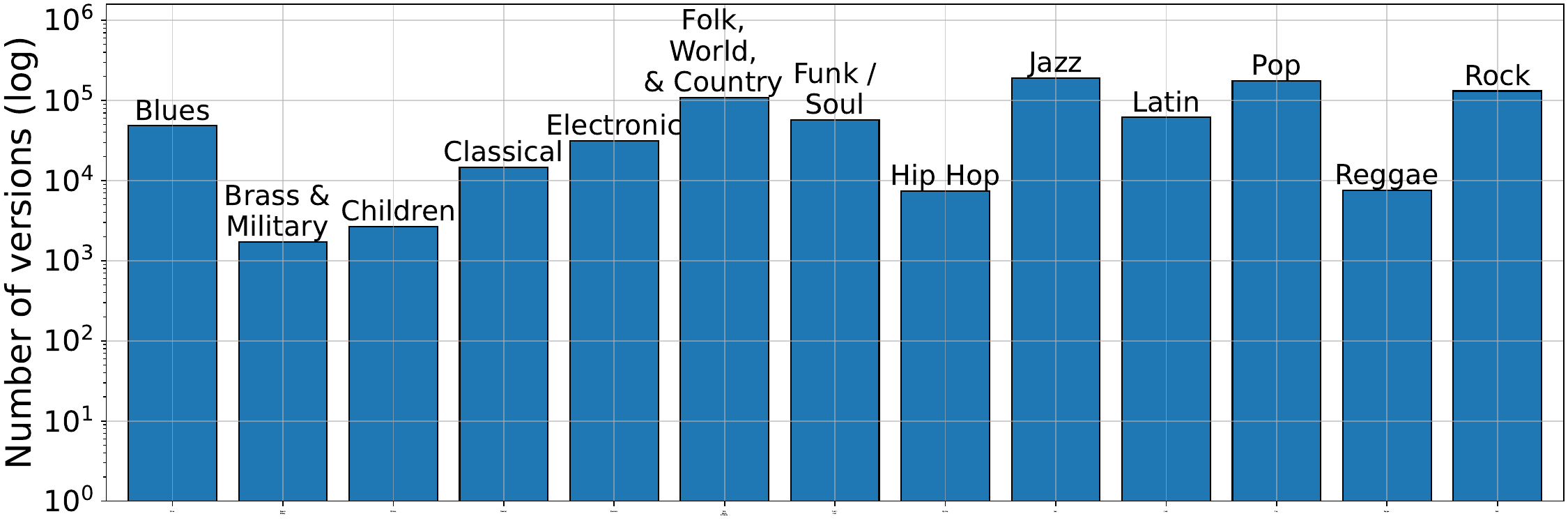}
    \caption{\ytdataset{} version genre distribution.}
    \label{fig:genre}
\end{figure}

Figure~\ref{fig:genre} reports the genre distribution of \ytdataset{}, demonstrating significant coverage over 13 genres. The distribution of styles, which is included in the project repository, covers 512 styles from Mambo to Tech House. Importantly, such genre metadata opens new possibilities for developing and evaluating VI systems. Previous studies have not delved into genre and style analyses, leaving their effect on performance underexplored. Given that our dataset contains relatively reliable genre and style annotations\footnote{Discogs genre and style annotations are release-level, however, they serve as a reasonable approximation for individual tracks.} such analysis is now possible \cite{bogdanov_acousticbrainz_2019}.\par

Table~\ref{tab:artists} compares the total number of artists of several VI datasets. Da-TACOS and SHS100K datasets provide only one artist per version while \dataset{} offers multiple. For a consistent comparison, we count one artist per \dataset{} version and do not include the group members. In addition, Da-TACOS noise works are not considered. The number of versions and artists comparisons between SHS100K and \ytdataset{} implies that our dataset contains more versions per artist on average.\par

\begin{table}[ht]
    \centering
    \small
    \begin{tabular}{lr}
        \toprule
            Dataset & Artists \\
        \midrule
            Da-TACOS & 6,375 \\
            SHS100K & 34,170 \\
        \midrule
            \ytdataset{} & 67,345 \\
            \dataset{} & 239,949 \\
        \bottomrule
    \end{tabular}
    \caption{Number of track artist comparison between selected datasets. One artist per version is reported.}
    \label{tab:artists}
\end{table}

\begin{figure}[ht]
    \centering
    \includegraphics[width=\columnwidth]{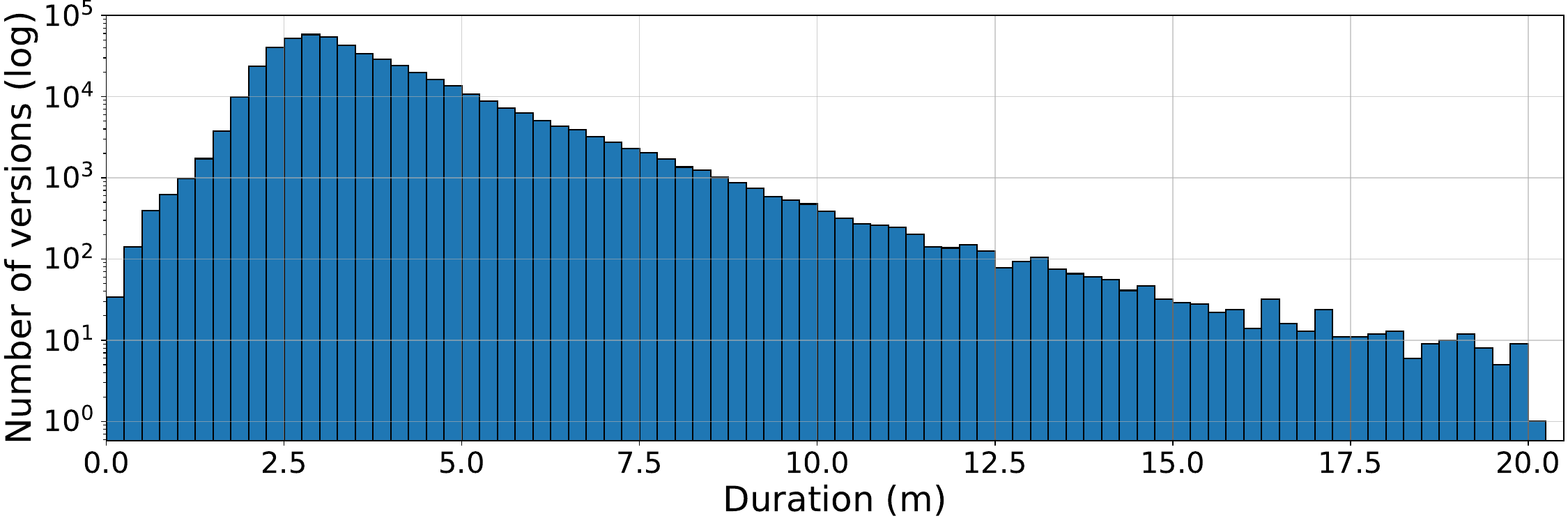}
    \caption{\ytdataset{} audio duration distribution.}
    \label{fig:durations}
\end{figure}

Figure~\ref{fig:durations} reports the audio duration distribution of \ytdataset{}, reflecting a comprehensive music collection. We observed that the long-duration tracks are mostly live versions and jazz or electronic music tracks, which can be notoriously long. Having long tracks increases the difficulty of training VI systems due to requiring effective time aggregation techniques or small embedding dimensions.\par

\subsection{Development and test splits}\label{sec:dataset:splits}
We split the \ytdataset{} dataset into training, validation, and test sets. To increase the compatibility with other datasets, the cliques in \dataset{} that intersect with the Da-TACOS benchmark and SHS100K-Test sets are ensured to be part of our test set. We excluded CoversDataset from this consideration due to its lack of metadata.\par

To determine the intersection between our dataset and the Da-TACOS benchmark set, we conducted a thorough comparison of track titles and track writers using artist names, aliases, and name variations. We successfully identified 935 out of the 1,000 (93\%) Da-TACOS cliques and 1,412 out of the 2,000 (71\%) ``noise'' tracks. Given the detailed artist metadata we employed, it is unlikely that the unidentified works are included in our training set. Moreover, since Da-TACOS selects its ``noise'' tracks from those lacking alternate versions and our \dataset{} consists exclusively of tracks with at least two versions, these tracks are also unlikely to be included in our training set. Regarding the SHS100K-Test set, we identified 1,555 out of the 1,692 cliques (90\%). The union of the identified cliques from both datasets is reserved for our test set.\par

\begin{figure}[!ht]
    \centering
    \includegraphics[width=\columnwidth]{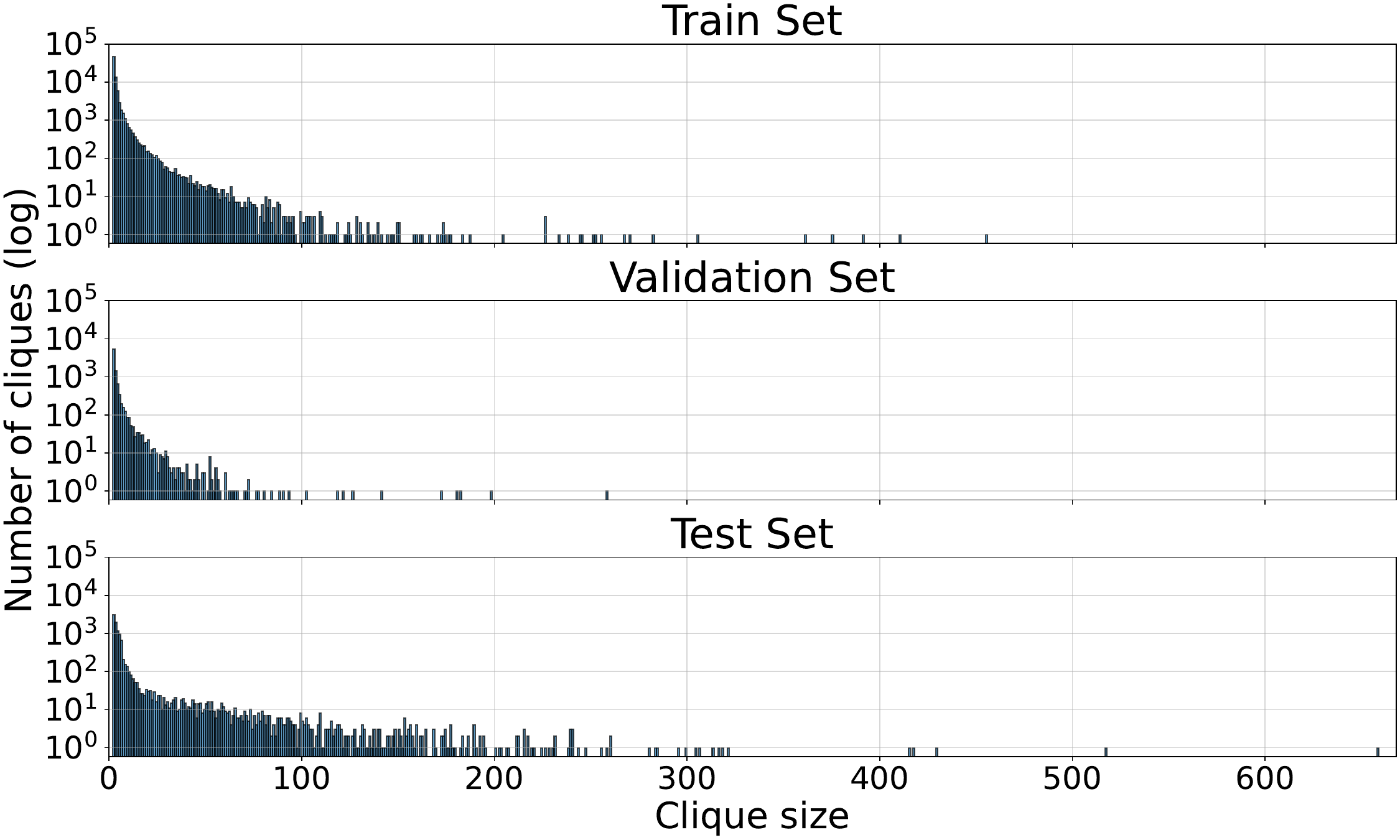}
    \caption{\ytdataset{} splits clique size distributions.}
    \label{fig:splits}
\end{figure}

We aimed for a 90-10\% development-test split; therefore, we sampled new cliques to add to the reserved cliques. While sampling the additional cliques, we did not exclude the SHS100K-Train set to use our dataset without restrictions. The reserved cliques from the Da-TACOS benchmark and SHS100K-Test sets had large enough sizes in our dataset. Moreover, similar to \cite{doras_prototypical_2020}, we believe that having small-sized cliques in the test set simulates real use cases better. Therefore, we randomly sampled the additional cliques from sizes two to six. The remaining cliques were assigned to the development set and were further partitioned into training and validation sets following a 90-10\% split. Figure~\ref{fig:splits} shows the clique size distribution of our splits, and Table~\ref{tab:partitions} compares the split sizes of different datasets.\par

\begin{table}[!ht]
\setlength{\tabcolsep}{0.07cm}
    \centering
    \small
    \begin{tabular}{@{}llrrrrr@{}}
    \toprule
        Dataset & Split & Cliques & Versions & MCS & ACS & mCS \\
    \midrule
        \multirow{2}{*}{Da-TACOS} & Benchmark & 1,000 & 13,000 & 13 & 13 & 13 \\
     & Noise & 2,000 & 2,000 & - & - & - \\
    \midrule
        \multirow{3}{*}{SHS100K} & Test & 1,692 & 10,547 & 162 & 6 & 5 \\
        & Validation & 1,842 & 10,884 & 17 & 6 & 6 \\
        & Train & 5,324 & 87,091 & 359 & 16 & 12 \\
    \midrule
        \multirow{3}{*}{\ytdataset{}} & Test & 9,878 & 116,197 & 658 & 12 & 3 \\
         & Validation & 8,890 & 37,081 & 258 & 4 & 2 \\
         & Train & 80,017 & 339,771 & 455 & 4 & 2 \\
    \bottomrule
    \end{tabular}
    \caption{Dataset partition sizes. MCS: maximum clique size; ACS: average clique size; mCS: median clique size}
    \label{tab:partitions}
\end{table}

\subsection{Audio representations}\label{sec:subsets:representations}
We computed the following audio representations commonly used in VI systems: chroma, HPCP~\cite{serra_audio_2010}, and CQT~\cite{schorkhuber_constant-q_2010}. They are available under request for non-commercial scientific research purposes.\par

\begin{table*}[htbp]
\centering
\small
\begin{tabular}{llrcccccc}
    \toprule
        \multirow{2}{*}{Training data} & \multirow{2}{*}{Model} & \multirow{2}{*}{d} & \multicolumn{2}{c}{Da-TACOS} & \multicolumn{2}{c}{SHS100K-Test} & \multicolumn{2}{c}{SHS100K-Test\textsuperscript{**}} \\
        \cmidrule(lr){4-5} \cmidrule(lr){6-7} \cmidrule(lr){8-9}
         & & & MAP $\uparrow$ & MR1 $\downarrow$ & MAP $\uparrow$ & MR1 $\downarrow$ & MAP $\uparrow$ & MR1 $\downarrow$ \\
    \midrule
        \multirow{3}{*}{Da-TACOS} & MOVE\cite{yesiler_accurate_2020} & 4,000 & 0.495 & 48\textsuperscript{\textdagger} & \xmark & \xmark & \xmark & \xmark \\
        & MOVE\cite{yesiler_accurate_2020} & 16,000 & 0.507 & 46\textsuperscript{\textdagger} & \xmark & \xmark & \xmark & \xmark \\
        & Re-MOVE\cite{yesiler_less_2020} & 256 & 0.524 & 43\textsuperscript{\textdagger} & \xmark & \xmark & \xmark & \xmark \\
    \midrule
        \multirow{5}{*}{SHS100K-Train} & TTP-Net\cite{yu_temporal_2019} & 300 & \xmark & \xmark &  0.465 & 72 & \xmark & \xmark \\
        & CQT-Net\cite{yu_learning_2020} & 300 & \xmark & \xmark & 0.655 & 55 & \xmark & \xmark \\
        & ByteCover\cite{du_bytecover_2021} & 2,048 & \xmark & \xmark & 0.836 & 47 & \xmark & \xmark \\
        & ByteCover2\cite{du_bytecover2_2022} & 128 & \xmark & \xmark & 0.839 & 46 & \xmark & \xmark \\
        & ByteCover2\cite{du_bytecover2_2022} & 1,536 & \xmark & \xmark & 0.863 & 39 & \xmark & \xmark \\
    \midrule
        \multirow{3}{*}{SHS100K-Train\textsuperscript{*}} & ByteCover\cite{du_bytecover_2021} & 2,048 & 0.714 & 23 & \xmark & \xmark & \xmark & \xmark \\
        & ByteCover2\cite{du_bytecover2_2022} & 128 & 0.718 & 23 & \xmark & \xmark & \xmark & \xmark \\
        & ByteCover2\cite{du_bytecover2_2022} & 1,536 & 0.791 & 19 & \xmark & \xmark & \xmark & \xmark \\
    \midrule
        SHS100K-Train\textsuperscript{**} & LyraC-Net\cite{hu_wideresnet_2022} & 1,024 & \xmark & \xmark & \xmark & \xmark & 0.765 & 48 \\
    \midrule
        Private & LyraC-Net\cite{hu_wideresnet_2022} & 1,024 & 0.813 & 15 & \xmark & \xmark & 0.884 & 33 \\
    \midrule
        \ytdataset{} & \model{} & 512 & 0.607 & 24 & \xmark & \xmark & 0.660 & 61 \\
    \bottomrule
\end{tabular}
\caption{Performance comparison on the Da-TACOS benchmark and SHS100K-Test sets.  \textsuperscript{*} denotes that the Da-TACOS benchmark set tracks were removed, \textsuperscript{**} denotes that the corresponding authors of that model downloaded the available URLs (therefore LyraC-Net~\cite{hu_wideresnet_2022} and \model{} are not evaluated on the same data), d denotes the embedding dimension, \xmark \ denotes that the result was not available, and \textsuperscript{\textdagger} denotes the corrected calculations described in Section~\ref{sec:baseline_model:results_public}.} 
\label{tab:full_results}
\end{table*}

\section{Baseline Model}\label{sec:baseline_model}
To demonstrate the utility of \ytdataset{} we search for a baseline model that uses computationally inexpensive input representations and is feasible for training on a consumer-grade GPU.\par

TPP-Net~\cite{yu_temporal_2019} and its successor CQT-Net~\cite{yu_learning_2020} rely on the classification loss for training. Due to the large number of cliques in \ytdataset{}, these models are difficult to train on this dataset without modifications. ByteCover~\cite{du_bytecover_2021}, ByteCover2~\cite{du_bytecover2_2022}, and LyraC-Net~\cite{hu_wideresnet_2022} are also difficult to train as they employ the classification loss with additional losses and feature complex architectures having significantly more parameters. Additionally, the code and pre-trained weights for these three models are not publicly available. We do not consider ByteCover3~\cite{du_bytecover3_2023} and CoverHunter~\cite{liu_coverhunter_2023} as they do not target full-track inputs. MOVE~\cite{yesiler_accurate_2020} and Re-MOVE~\cite{yesiler_less_2020} are not considered due to their reliance on computationally expensive input representations. Ultimately, we selected CQT-Net, primarily due to its adaptability for use with \ytdataset{}.\par

\subsection{CQT-Net}\label{sec:baseline_model:cqtnet}
The original model is trained with the classification task, where clique IDs of the SHS100K dataset are used as class labels. A multi-length training strategy that presents the model with three different segments from each version is used to reduce possible biases toward input duration. Additionally, tempo change and spectral masking data augmentation techniques are used. During retrieval, the classification head is discarded and the remaining network is used for extracting version embeddings, whose similarity is computed with cosine similarity.\par

\subsection{\model{}}\label{sec:baseline_model:discogsvinet}
Training CQT-Net with classification loss is challenging due to the large number of cliques in \ytdataset{}. Therefore, we utilize the triplet loss, similar to previous research \cite{doras_cover_2019, yesiler_accurate_2020}. To this end, we remove the classification head from the architecture and change the affine projection layer to a linear projection with 512-dimensional outputs. Additionally, we include an $L_2$ normalization layer to ensure that embeddings lie on the unit hypersphere. The resulting model contains 5.2 million parameters.\par

At each training iteration, a mini-batch is created by randomly sampling 48 distinct cliques and two random versions per clique. With this configuration, each sample can only have one positive; hence, the positive mining strategy is equivalent to offline random sampling. For mining negatives, we use online hard-negative mining.\par

We extract the CQT input representations before training with CQT-Net's setting. However, we store them with 16-bit precision due to the large storage requirement of our dataset. Unlike CQT-Net's multi-length training strategy, we use fixed-length inputs where consecutive CQT frames of about 185 seconds are taken randomly. Then the features are mean downsampled with a factor of 20, following the authors. To demonstrate the benefits of using our large dataset, we do not use any data augmentation method during training, such as tempo and key modifications, spectral masking techniques, or audio degradation methods used in previous VI research.\par

We train \model{} for 50 epochs, which takes about 25 hours using a single Nvidia RTX2080. We use the AdamW optimizer, setting the initial learning rate to $1\text{e-}3$ and adjusting via exponential decay. The triplet loss margin is set to $0.1$.\par

During training, we use our validation set to monitor performance. Every five epochs, we simulate the VI task and save the best model in terms of mean average precision (MAP). However, we evaluate the model at the end of the training on \ytdataset{}, Da-TACOS, and SHS100K datasets using MAP and the mean rank of the first relevant item (MR1) metrics.\par

\subsection{Evaluation on \ytdataset{}}\label{sec:baseline_model:ours}
Due to potential overlaps between the training sets of publicly available VI models and the \ytdataset{} test set, we could not benchmark the publicly available models. For instance, as discussed in Section~\ref{sec:dataset:splits}, there can be shared tracks with the SHS100K-Train set. Similarly, the Da-TACOS training set, which is not publicly available, may share tracks with our test set, rendering comparisons with models trained on this dataset unreliable. Additionally, as discussed in Section~\ref{sec:baseline_model}, training numerous models on \ytdataset{} were not possible. We acknowledge these limitations and suggest that benchmarking models is a critical area for future research.\par

Despite these challenges, we present the scores obtained by \model{}. Our model obtains a MAP score of 0.443 and an MR1 score of 614.1 on the \ytdataset{} test set, which establishes the baseline scores on this dataset. The contrast between the MR1 and MAP values can be attributed to the realistic clique size distribution. As shown in Figure~\ref{fig:splits}, the test set contains numerous cliques with size two. When a query is made with a version from these cliques, retrieving the only other version in high rankings contributes significantly to the MAP metric.\par

\subsection{Evaluation on Da-TACOS and SHS100K}\label{sec:baseline_model:results_public}
We tested \model{} on the Da-TACOS benchmark and SHS100K-Test sets. From the SHS100K-Test set, we could download 8,489 versions (80\% of the total). As discussed in Section~\ref{sec:dataset:splits}, we perform an extensive analysis to ensure that our training set has a minimal intersection with the evaluated sets.\par

The results are presented in Table~\ref{tab:full_results}, relying on results reported in the literature except for MOVE and Re-MOVE, for which we recomputed the results due to a metric calculation problem we discovered. In the public Da-TACOS evaluation script, "noise" works are wrongly boosting the MR1 score instead of being excluded. We corrected this issue, tested the official MOVE and Re-MOVE models, and listed the updated MR1 values.\par

In Table~\ref{tab:full_results}, \model{} outperforms both MOVE and Re-MOVE on the Da-TACOS benchmark set, which is a significant improvement given the simplicity of our input representation and lack of data augmentations. Unlike such, \model{} does not depend on pre-trained models for input representation. As a result, it exhibits significantly faster embedding extraction, similar to those reported in \cite{du_bytecover2_2022}.\par

On the SHS100K-Test set, even though we used a slightly smaller subset due to some URLs becoming unavailable, we could not improve over other considered models, except for the TTP-Net and CQT-Net. In particular, CQT-Net, which we modified for our baseline, performed similarly. We posit that these differences may stem from the absence of data augmentation techniques in our methodology or from the classification loss possibly structuring the latent space more effectively than the triplet loss we implemented. Nonetheless, further experiments are required.\par

ByteCover, ByteCover2, and LyraC-Net outperform \model{} by a significant margin. This performance difference can be attributed to several factors: the combined use of classification and triplet losses, as reported in the literature \cite{luo_bag_2019}, the advantages obtained by training larger architectures, or the absence of data augmentations in our model. However, it is important to note that independent studies have raised concerns about the reproducibility of the published results associated with the ByteCover approach \cite{ohanlon_detecting_2021, hu_wideresnet_2022}.\par

\section{Conclusion}\label{sec:conclusion}
We presented a new methodology to create a VI dataset from a previously unused metadata source, Discogs. Using this metadata, we identified a large number of cliques and versions to create the \dataset{} dataset and matched a large portion of the versions with official YouTube URLs to create its \ytdataset{} subset. Our datasets surpass existing datasets by far in size and provide unprecedented metadata detailing genre, style, and artist relationships.\par

To demonstrate the utility of \ytdataset{}, we trained a baseline model, \model{}, on the training set and evaluated the model performance on the test set, establishing baseline results. Additionally, we assessed \model{}'s performance on the Da-TACOS benchmark and SHS100K-Test sets, where it demonstrated competitive performance. Notably, our model achieved these results without relying on any data augmentation techniques, multiple training losses, or complex architectural designs.\par

We leave training large models, using the metadata relations for training and evaluation, and investigating the role of data augmentations as future work.\par

\clearpage

\section{Acknowledgements}\label{sec:acknowledgements}
This work is supported by ``IA y Música: Cátedra en Inteligencia Artificial y Música’' (TSI-100929-2023-1) funded by the Secretaría de Estado de Digitalización e Inteligencia Artificial and the European Union-Next Generation EU, under the program Cátedras ENIA.\par

\bibliography{VersionIdentification}

\end{document}